\documentclass[aps,prl,amsmath,amsfonts,amssymb,twocolumn,showpacs]{revtex4}
\usepackage[latin1]{inputenc}
\usepackage{graphicx}
\usepackage[english]{babel}

\usepackage{amsfonts}
\usepackage{amsmath}
\usepackage{amssymb}

\newcommand{\beq}{\begin{equation}}
\newcommand{\eeq}{\end{equation}}
\newcommand{\nn}{\nonumber}
\newcommand{\ket}[1]{|#1\rangle}
\newcommand{\bra}[1]{\langle #1|}

\makeatletter
\@ifundefined{textcolor}{}
{%
 \definecolor{BLACK}{gray}{0}
 \definecolor{WHITE}{gray}{1}
 \definecolor{RED}{rgb}{1,0,0}
 \definecolor{GREEN}{rgb}{0,1,0}
 \definecolor{BLUE}{rgb}{0,0,1}
 \definecolor{CYAN}{cmyk}{1,0,0,0}
 \definecolor{MAGENTA}{cmyk}{0,1,0,0}
 \definecolor{YELLOW}{cmyk}{0,0,1,0}
 }

\begin{document}

\title{Dynamic Stark shift induced by a single photon packet}

\author{D. Valente
$^{1}$
}
\email{daniel@fisica.ufmt.br}

\author{F. Brito
$^{2}$
}

\author{T. Werlang
$^{1}$
}

\affiliation{
$^{1}$ 
Instituto de F\'isica, Universidade Federal de Mato Grosso, CEP 78060-900, Cuiab\'a, MT, Brazil
}

\affiliation{
$^{2}$ 
Instituto de F\'isica de S\~ao Carlos, Universidade de S\~ao Paulo, C.P. 369, 13560-970, S\~ao Carlos, SP, Brazil
}

\begin{abstract}
The dynamic Stark shift results from the interaction of an atom with the electromagnetic field.
We show how a propagating single-photon wave packet can induce a time-dependent dynamical Stark shift on a two-level system (TLS).
A non-perturbative fully-quantum treatment is employed, where the quantum dynamics of both the field and the TLS are analyzed.
We also provide the means to experimentally access such time-dependent frequency by measuring the interference pattern in the electromagnetic field inside a one-dimensional (1D) waveguide.
The effect we evidence here may find applications in the autonomous quantum control of quantum systems without classical external fields, that can be useful for quantum information processing as well as for quantum thermodynamical tasks.
\end{abstract}

\maketitle

The Stark shift is a variation of the energy levels of an atom under the influence of an electric field.
Originally found for a static field, the Stark effect has also been found for an atom driven by an alternating electric field \cite{stark}.
This is known as the Autler-Townes effect, which is called the dynamic or AC Stark shift as well.
Dynamic Stark shifts have additionally been found for the case in which an atom is perturbed by the scattering of a single monochromatic photon \cite{cohen}.

The groundbreaking experimental expertise in controlling the quantum dynamics of single photons interacting with single real or artificial atoms of the last decade offers unprecedented research avenues.
In particular, the advent of nano or micro one-dimensional (1D) waveguides, in the areas of semiconductor nanophotonics, of nano optical fibers and of superconducting circuits, are worth of special attention.
They allow efficient propagation of wave packets, whereas keeping light-matter coupling at the single photon level,
providing novel regimes as compared to cavity quantum electrodynamics or nonlinear photonic crystal scenarios.
For instance, a single quantum dot inside a 1D waveguide can form an efficient single photon source \cite{jmg,lodahl08}. 
A single-photon wave packet can then be routed through the 1D waveguide \cite{wallraff} until it reaches another atom, with which it can interact \cite{sand, ck, astafiev, kimble14, KimblePRL15, faez}, even inducing non-Markovian dynamics in that atom \cite{OL}.
Single atoms coupled to 1D waveguides form, thus, a strong candidate for quantum communication and quantum information processing purposes \cite{chang07, sand09,tsai10,dv12,fratini}.

In this work, we characterize the dynamic Stark shift induced by a propagating single-photon wave packet on a two-level system (TLS).
We employ a non-perturbative, analytic, fully-quantum treatment, where the quantum dynamics of both the field and the TLS are taken into account.
We also evidence how that time-dependent frequency shift can be measured from the interference pattern in the field coming from the waveguide.


\begin{figure}[!htb]
\centering
\includegraphics[width=1\linewidth]{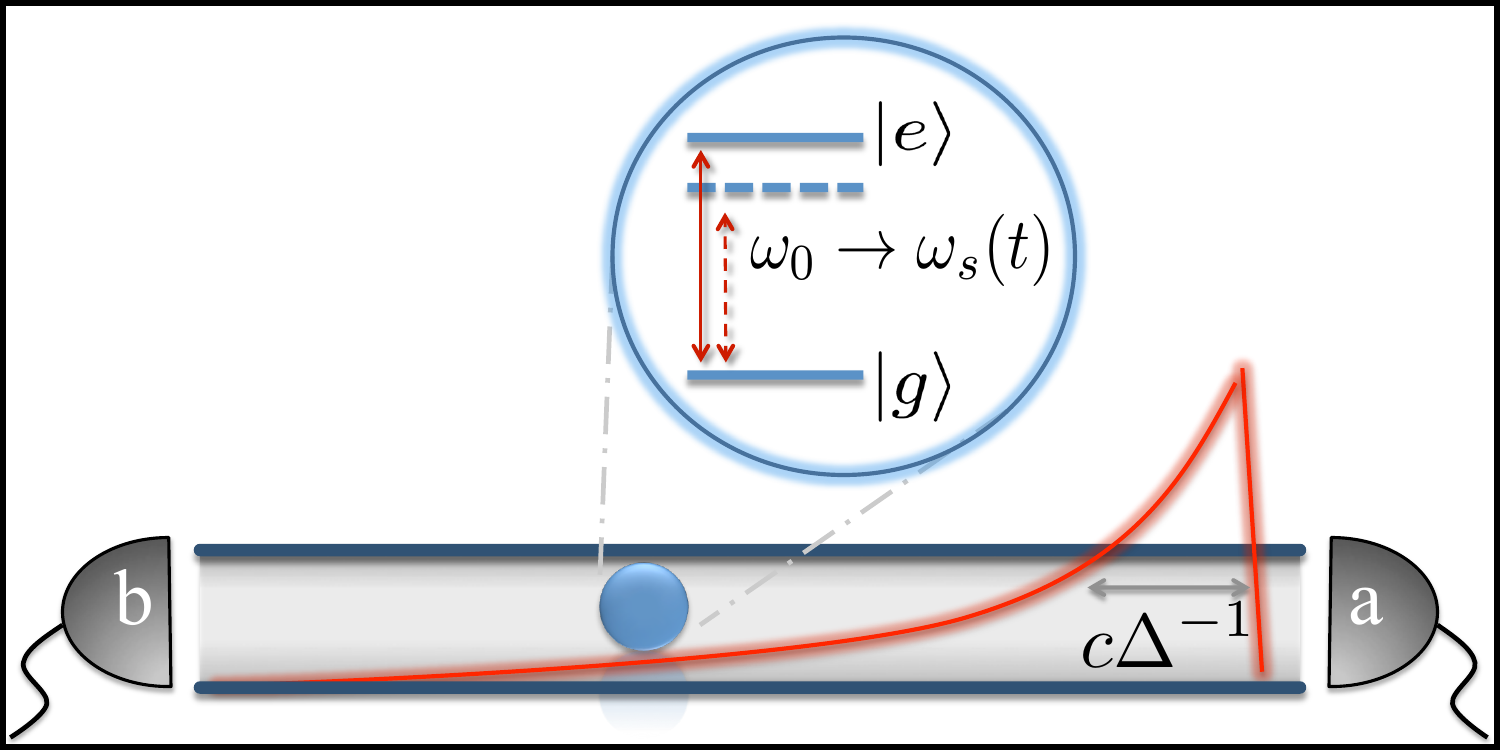}
\caption{(Color online)
{\bf Dynamic Stark shift of the TLS.}
A two-level system (sketched in the blue full circle), described by states $\ket{g}$ and $\ket{e}$, is placed inside a one-dimensional waveguide.
A right-propagating single-photon packet (exponential-shaped red line) is sent through the waveguide to interact with the TLS. 
The spatial spread of the photon packet is $c \Delta^{-1}$, where $c$ is the speed of light inside the waveguide and $\Delta$ is the spectral linewidth of the pulse.
During its interaction time with the TLS, the photon packet modulates the TLS transition frequency in time, $\omega_0 \rightarrow \omega_s(t)$.
After the photon has interacted with the atom it can be reemitted forward (through channel $a$) or backwards (channel $b$), being measured by the respective photodetectors.
Both the dynamics of the TLS and the electromagnetic field are addressed by means of a non-perturbative fully-quantum approach.
} 
\label{model}
\end{figure}




The dynamics of the composite system is unitary, governed by the total Hamiltonian 
$H = H_{\mathrm{system}}+H_{\mathrm{int}}+H_{\mathrm{field}}$.
Here,  
$H_{\mathrm{system}} = \hbar \omega_{\mathrm{isol}} \sigma_{+}\sigma_{-}$
is the Hamiltonian of the isolated TLS, in which $\sigma_{-} = \sigma_{+}^\dagger = \ket{g}\bra{e}$.
The TLS ground (resp. excited) state is denoted by $\ket{g}$ (resp. $\ket{e}$), as sketched in Fig.\ref{model}.
The Hamiltonian of the 1D free field modes that propagate forwards $a_\omega$ and backwards $b_\omega$, with frequency $\omega$, inside the waveguide reads
$H_{\mathrm{field}} = \sum_{\omega} \hbar \omega [a^\dagger_\omega a_\omega + b^\dagger_\omega b_\omega] $.
The TLS-field interaction Hamiltonian, in the dipole and rotating-wave approximations, is given by 
\cite{domokos}
$H_{\mathrm{int}} = 
\sum_\omega -i \hbar g_\omega 
[\sigma_{+} (a_\omega e^{+i k_\omega x_s}+ b_\omega e^{-i k_\omega x_s}) - \mbox{H.c.}]$,
in which $x_s$ is the position of the TLS, 
$k_\omega = \omega/c$ is the wavevector modulus, $c$ is the speed of light inside the waveguide, $g_\omega$ is the coupling rate and H.c. denotes the Hermitian conjugate.
Ref.\cite{domokos} also provides a detailed analysis of the waveguide mode area $\mathcal{A}$, i.e. its effective transverse cross section, and its influence on $g_\omega\propto 1/\sqrt{\mathcal{A}}$.
The reference frame is set so that $x_s = 0$.
We are interested in the zero- and one-excitation subspace, as described by the normalized pure state of the global TLS-plus-field system, 
$\ket{\xi(t)} = c_0\ket{g,0} + \psi(t) \ket{e,0} +
\sum_\omega [\phi^{(a)}_\omega(t) a^\dagger_\omega + \phi^{(b)}_\omega(t) b^\dagger_\omega ] \ket{g,0}$,
where $\ket{0}$ is the vacuum state of the field.
The excited-state population of the TLS is $|\psi(t)|^2$.
The amplitude $c_0$ is static and we choose $c_0 = 0$.
The real-space representation of the field reads \cite{OL} 
$\phi^{(a)}(x,t) = \sum_{\nu} \phi^{(a)}_{\nu}(t) e^{ik_{\nu}x}$ for the $a_\nu$ modes.
It represents the probability amplitude that the photon is found at position $x$, at time $t$, propagating forwards.
For the  $b_\nu$ modes, one substitutes $k_\nu$ for $-k_\nu$ as the photon propagates backwards.
A continuum of frequencies is assumed for the 1D environment, 
$\sum_\nu \rightarrow \int d\nu \rho_{\mathrm{1D}}$, 
so the flat spectral density of guided modes is named $\rho_{\mathrm{1D}}$.

We analytically solve the Schr\"odinger equation $i\hbar \partial_t \ket{\xi(t)} = H\ket{\xi(t)}$ to find the dynamics of the composite TLS-plus-field system.
Because the hamiltonian $H$ conserves the number of excitations, $\ket{\xi(t)}$ is still restricted to the single-excitation subspace.
The self interaction of the TLS mediated by the electromagnetic reservoir imposes a Lamb shift,
$\omega_{\mathrm{Lamb}} = \sum_\nu g_\nu^2 \mathcal{P}\left\{1/(\omega_{\mathrm{isol}}-\nu) \right\}$, 
where $\mathcal{P}$ stands for the principal part.
So, the effective frequency of the TLS becomes $\omega_0 = \omega_{\mathrm{isol}} + \omega_{\mathrm{Lamb}}$.
The continuum of frequency modes imposes a vacuum-induced decay rate to the TLS, 
$\Gamma_{\mathrm{1D}} = 4\pi g_{\omega_0}^2 \rho_{\mathrm{1D}}$, 
in the Wigner-Weisskopf approximation.
For a general input photon packet the excited-state amplitude dynamics is found to be
$\psi(t) = \psi(0) e^{ -\left( \frac{\Gamma_{\mathrm{1D}}}{2}+i\omega_0 \right) t }
-\sqrt{\frac{\Gamma_{\mathrm{1D}}}{4\pi\rho_{\mathrm{1D}}}} 
\int_0^t \left[ \phi^{(a)}(-ct',0) + \phi^{(b)}(ct',0) \right] e^{-\left(\frac{\Gamma_{\mathrm{1D}}}{2} + i\omega_0 \right)(t-t')} dt'$.
Hence, the excited-state amplitude is driven by the value of the initial photon packet at the position of the atom. 
This represents a state of the field that is out of thermal equilibrium.


We now analyze the reduced TLS dynamics. 
Our aim is to identify the dynamic shift of the TLS frequency, $\omega_0 \rightarrow \omega_s(t)$, induced by the out-of-equilibrium quantum state of the electromagnetic environment, represented by a propagating single-photon packet.
That is evidenced by a non-perturbative treatment, where the reduced dynamics of the TLS is rewritten in the form of an exact master equation, which happens to be in a Lindblad form.
Such reduced dynamics of the TLS is directly obtained by tracing out the field variables, 
$\rho_s(t) = \mbox{Tr}_{\mathrm{field}}\{ \ket{\xi(t)} \bra{\xi(t)} \}$.
In close analogy to the procedure of Ref.\cite{breuer07}, we show that $\rho_s(t)$ obeys the following master equation,
\beq
\partial_t \rho_s(t) = -\frac{i}{\hbar} [H_s(t),\rho_s(t)] + \mathcal{L}_t \{ \rho_s(t) \},
\label{ME}
\eeq
where 
\beq
H_s(t) = \hbar \omega_s(t) \ \sigma_+ \sigma_-
\label{MEH}
\eeq
and
$
\mathcal{L}_t \{ \rho_s(t) \} 
= 
\Gamma(t) \left( \sigma_- \rho_s(t) \sigma_+ - \frac{1}{2} \{  \sigma_+ \sigma_-, \rho_s(t) \}  \right),
$
$\{.,.\} $ being the anticommutator.
In Eq.(\ref{MEH}), the TLS time-dependent frequency induced by the single-photon packet is defined by 
\beq
\omega_s(t) \equiv -\mbox{Im}[\partial_t \psi(t)/\psi(t)]
\label{defomegas}
\eeq
and the time-dependent decay rate induced by the quantized field is defined by
$\Gamma(t) \equiv  -2\ \mbox{Re}[\partial_t \psi(t)/\psi(t)]$.
These expressions provide the analytical means by which one can obtain both the unitary and non-unitary parts of the TLS reduced dynamics.
The former is driven by the effective TLS Hamiltonian $H_s(t)$ and the latter, by the effective Lindbladian $\mathcal{L}_t$.
Both terms have the same physical origin, namely, the interaction of the TLS with the 1D continuum of modes of the electromagnetic field, in which a single-photon packet state can be initially prepared.

The time-dependent TLS frequency evidenced by the master equation (\ref{ME}) has a sound physical meaning.
A propagating photon pulse represents out-of-equilibrium fluctuations of the electric field amplitude in time and space.
Since the TLS-field interaction $H_{\mathrm{int}}$ is given by means of a dipole coupling, a change in the electric field may stretch and contract the TLS dipole, inducing the time-dependent Stark shift effect.
To make this statement more precise, we can show that
$\partial_t {\omega}_s(t) = (1/2)\ \partial_t \left( \langle H_{\mathrm{int}} (t) \rangle / |\psi(t)|^2 \right)$,
where 
$\langle H_{\mathrm{int}} (t) \rangle = \bra{\xi(t)} H_{\mathrm{int}} \ket{\xi(t)}$.
So, the time variation of the TLS frequency is related to the time variation of the average dipole interaction energy, 
as intuitively expected.
In particular, if the average interaction energy is zero, the TLS frequency becomes static, $\partial_t{\omega}_s(t) = 0$.
We have found that
$\langle H_{\mathrm{int}} (t) \rangle = 2\hbar g_{\omega_0} \mbox{Im}[\phi^{(a)}(0,t) \psi^*(t)]$,
which vanishes if the phase of the field amplitude is equal to the phase of the TLS excited-state amplitude.

\begin{figure}[!htb]
\centering
\includegraphics[width=1\linewidth]{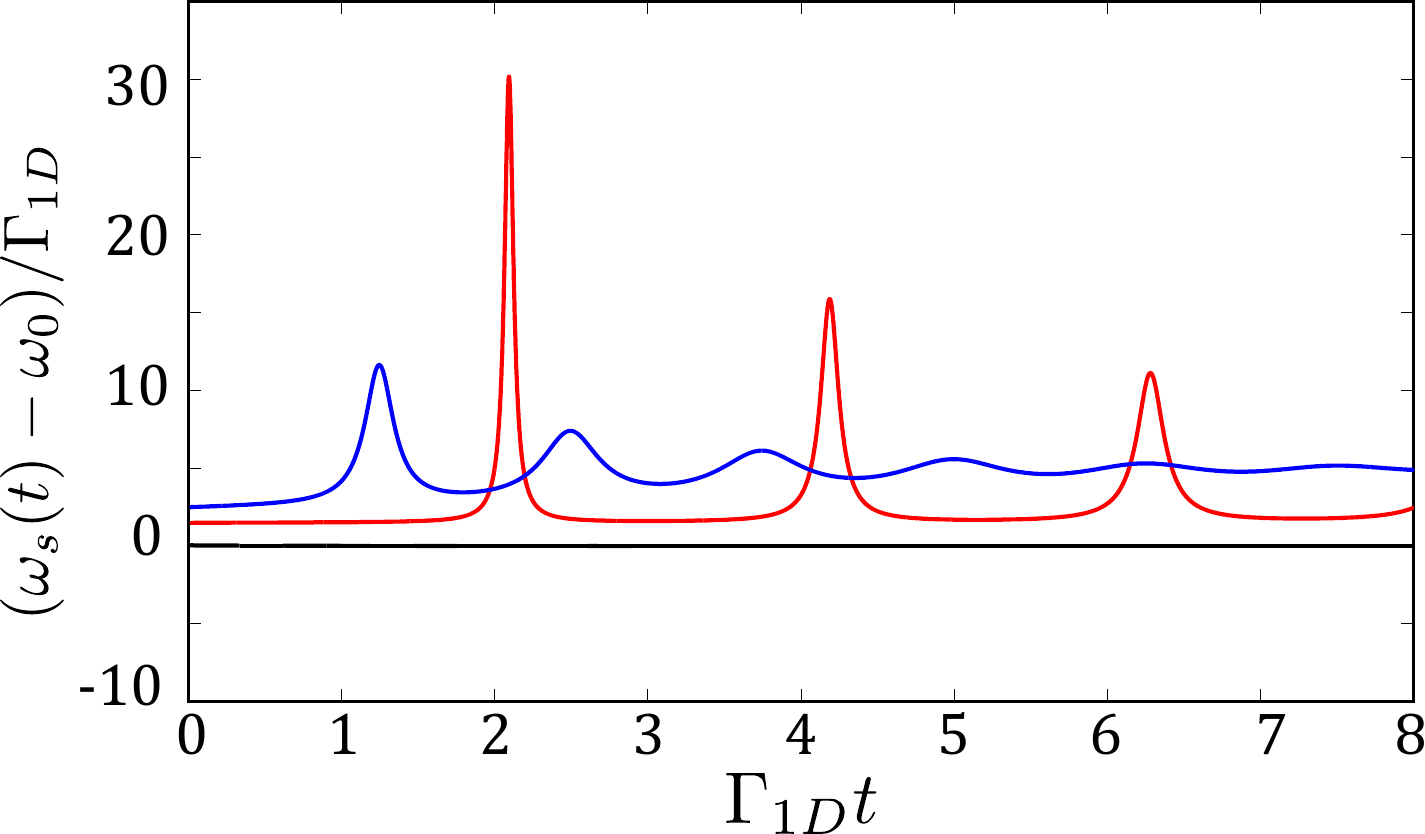}
\caption{
(Color online) {\bf Time-dependent TLS frequency.}
The blue curve stands for a highly detuned $\delta = 5$ and long packet (small linewidth) $\Delta = 0.1$.
The black curve stands for a low detuning $\delta = 0.1$ and short packet (large linewidth) $\Delta = 5$.
The red curve stands for intermediate detuning $\delta = 3$ and linewidth $\Delta = 0.9$.
In all the cases, we have chosen 
$\omega_0 = 10^6\ \Gamma_{\mathrm{1D}}$ and $\Gamma_{\mathrm{1D}} = 1$.
In the case of realistic quantum dots in photonic nanowires \cite{jmg}, 
 $\Gamma_{\mathrm{1D}} \sim 10^2$ MHz and $\omega_0 \sim 3 \times 10^6 \Gamma_{\mathrm{1D}} = 300$ THz. 
In such more realistic scenario the shift can be of the order of $\omega_s(t) - \omega_0 \sim 10^{-3}\% \times  \omega_0 = 10\  \Gamma_{\mathrm{1D}} \sim 10^3$ MHz.
By repeating the analyzes for the parameters of Refs.\cite{sand,faez,bleuse,arcari,kimble14}, we always find $|\omega_s(t) - \omega_0| \sim  10\  \Gamma_{\mathrm{1D}} $, in each case.} 
\label{w}
\end{figure}

An explicit expression for $\omega_s(t)$ has been found for the particular case where the incident photon packet is prepared by means of a spontaneous emission from another TLS, of frequency $\omega_L$ and linewidth $\Delta$. 
In that case, the field amplitude has an exponential profile \cite{sand, OL},
$\phi^{(a)}(x,0) = N \Theta(-x) \exp[(\Delta/2 + i\omega_L)x/c]$, where 
$N = \sqrt{2\pi \rho_{\mathrm{1D}}\Delta}$
is a normalization factor and $\Theta(x)$ is the Heaviside step function.
The packet spectral linewidth is characterized by $\Delta$, hence the typical time duration of the pulse is $\Delta^{-1}$.
The central frequency of the packet is $\omega_L$, that corresponds to the transition frequency of the emitter.
We define a static detuning $\delta = \omega_L - \omega_0$.
In addition, we choose $\psi(0)=0$ and $\phi^{(b)}(x,0)=0$ to find
$\psi(t) = -\sqrt{\frac{\Gamma_{\mathrm{1D}} \Delta}{2}}  e^{-\left( \frac{\Gamma_{\mathrm{1D}}}{2} + i\omega_0 \right) t }
\left[ \frac{e^{\left( \frac{\Gamma_{\mathrm{1D}} - \Delta}{2}- i\delta \right)t }- 1 }{\frac{\Gamma_{\mathrm{1D}} - \Delta}{2}- i\delta}\right]$. 
In that case, the time-dependent frequency reads
\beq
\omega_s(t) = \omega_0 +  
        \partial_t \tan^{-1} 
          \left( 
\frac{\sin(\delta t)}{\cos(\delta t)  - e^{(\Delta - \Gamma_{\mathrm{1D}} ) t/2 } }
         \right).
\label{wparticular}
\eeq
The function $\omega_s(t)$ is plotted in Fig.\ref{w}. 
Three regimes are presented.
The blue curve stands for a highly detuned $\delta = 5$ and long packet (small linewidth) $\Delta = 0.1$ 
(in $\Gamma_{\mathrm{1D}} = 1$ units).
We see highly oscillatory frequency shifts.
The black curve stands for a low detuning $\delta = 0.1$ and short packet (large linewidth) $\Delta = 5$.
The small duration of the pulse does not leave time enough for any shift to be seen.
In addition, the almost resonant photon cannot induce a significant shift in the TLS transition.
The red curve stands for intermediate detuning $\delta = 3$ and linewidth $\Delta = 0.9$.
There we see the highest peaks in the dynamic shift.
In all the cases, we have chosen $\omega_0 = 10^6 \Gamma_{\mathrm{1D}}$. 
Eq.(\ref{wparticular}) shows that the curve shape does not depend on $\omega_0$, apart from a multiplicative factor.
For realistic quantum dots in photonic nanowires \cite{jmg} the order of magnitude of the shift is $\sim 10^{-3}\% \times \omega_0 = 10^{-3}\% \times 10^6 \Gamma_{\mathrm{1D}} = 10\  \Gamma_{\mathrm{1D}} \sim 10^3$ MHz, since 
$\Gamma_{\mathrm{1D}} \sim 10^2$ MHz and $\omega_0 \sim 3 \times 10^6 \Gamma_{\mathrm{1D}} = 300$ THz.
In other words the shift, $\omega_s(t)-\omega_0$, can be $10$ times bigger than the transition line resolution, $\Gamma_{\mathrm{1D}}$.
The parameters of Refs.\cite{sand,faez,bleuse,arcari,kimble14} all give rise to the same order of magnitude for the shifts, i.e., $\omega_s(t) - \omega_0 \sim  10\  \Gamma_{\mathrm{1D}} $.
Analytically, from Eq.(\ref{wparticular}), we see that ${\omega}_s(t) - \omega_0$ is an odd function of the detuning $\delta$.
Thus, at zero detuning, $\delta=0$, the time-dependent frequency becomes ${\omega}_s(t) = \omega_0$.
Under the mode-matching condition, $\Delta = \Gamma_{\mathrm{1D}}$, the Stark shift also becomes time independent, ${\omega}_s(t) = \omega_0 + \delta/2$.
For a long packet ($\Delta  \ll  \Gamma_{\mathrm{1D}}$), the time dependency rapidly vanishes, so the photon also induces only a static frequency shift in the TLS,
$\omega_s(t) \rightarrow \omega_0 + \delta$.
Therefore, the TLS frequency becomes equal to the photon central frequency, $\omega_s(t) \approx \omega_0 + \delta = \omega_L$.
The intuitive explanation for this result in the long packet regime relies in the fact that the atom approximately behaves as a dielectric medium, in the sense that its dipole oscillates at the same frequency of the pumping field.
The atom dipole is proportional to the TLS excited-state amplitude, which becomes approximately
$\psi(t) \propto \exp[-(\Delta/2+i \omega_L )t ]$, making it clear that the central frequency of the TLS is $\omega_L$ in the 
$\Delta  \ll  \Gamma_{\mathrm{1D}}$ limit.
In this case, the outgoing photon will have the same frequency as the incoming one.
Note that, although the shift increases linearly with the detuning, its impact on the field observables decay as $\sim 1/(\Gamma^2 + 4 \delta^2)$, as discussed below Eq.(\ref{Interference}).
For the short packet ($\Delta \gg \Gamma_{\mathrm{1D}}$) regime, we see that, again, the time-dependent frequency becomes static, $\omega_s(t) \approx \omega_0$.

%
%
%

\begin{figure}[!htb]
\centering
\includegraphics[width=1\linewidth]{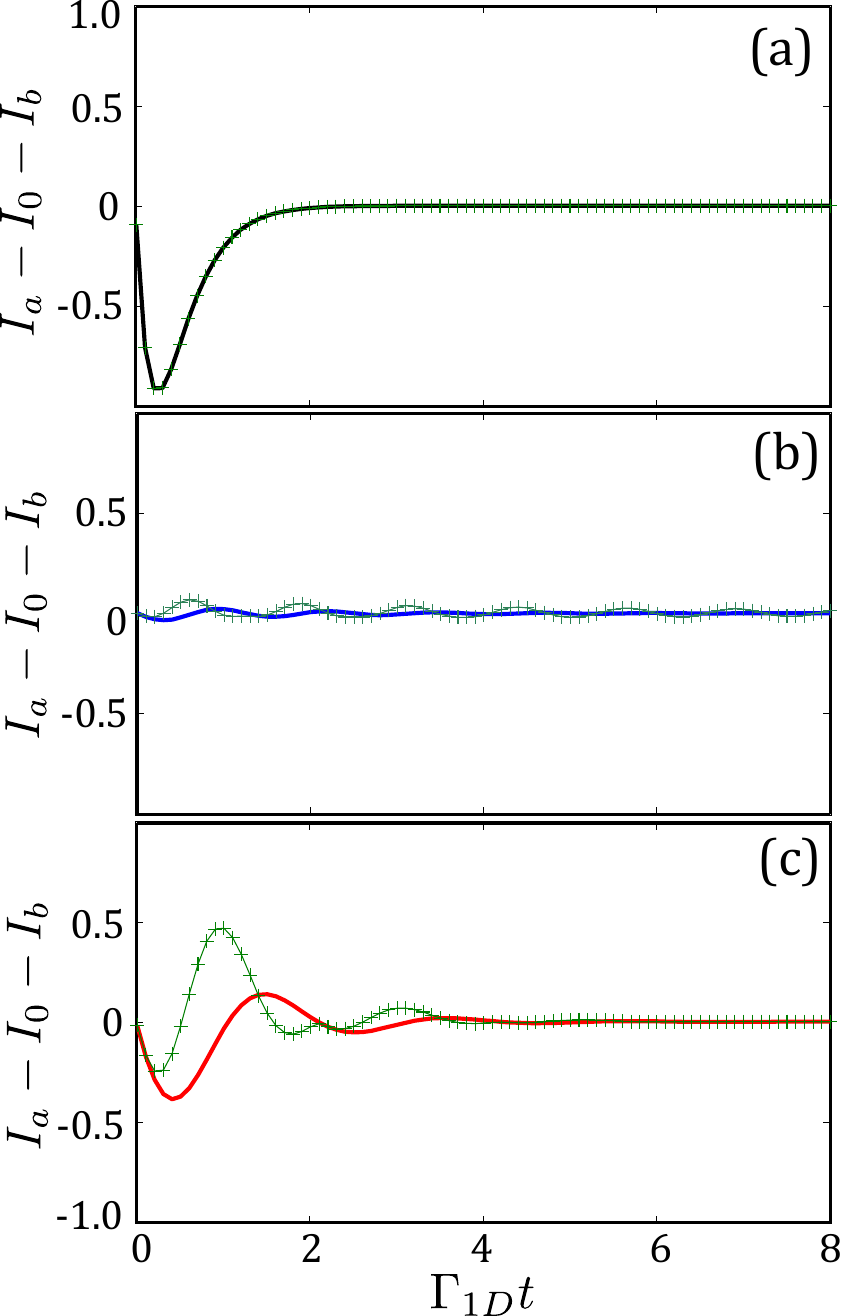}
\caption{(Color online).
{\bf Measuring $\omega_s(t)$ from field intensities.}
The difference of the signals $(I_a - I_0) - I_b$ is plotted as a function of time.
In all plots, the full lines represent the the case in which the correct time-dependent frequency $\omega_s(t)$ is taken into account, whereas the crosses represent the case where $\omega_s(t)$ is incorrectly approximated by the
static frequency $\omega_0$.
Plot (a) corresponds to the low detuned ($\delta = 0.1$) short packet ($\Delta = 5$).
Plot (b) corresponds to the highly detuned ($\delta = 5$) long packet ($\Delta = 0.1$).
Plot (c) corresponds to the intermediate detuning ($\delta = 3$) and linewidth ($\Delta = 0.9$), in which case the contrast is most visible.} 
\label{intensity}
\end{figure}
The output channels of the waveguide that establishes the 1D electromagnetic environment give information about the TLS dynamics, if one takes the perspective of seeing such TLS dynamics as a photon absorption-(re)emission process.
At channel $b$, for instance, the field amplitude $\phi^{(b)}(x_d,t)$ at the position of the detector $x_d < 0$ at time $t$ is directly proportional to the excited state amplitude, 
$\phi^{(b)}(x_d,t) = \sqrt{\beta} \ \psi(t-|x_d|/c)$, where
$\beta = \Gamma_{\mathrm{1D}} \pi \rho_{\mathrm{1D}}$.
The probability density of detecting a photon at position $x_d$ at time $t$, which gives the intensity $I_b(t)$ of the electric field at the detector, is 
$I_b(t) = |\phi^{(b)}(x_d,t) |^2 = \beta |\psi(t-|x_d|/c)|^2$.
The signature of the time dependency of $\omega_s(t)$ appears on the field $\phi^{(b)}(x_d,t)$.
By writing $\psi(t) = |\psi(t)|\exp{i\theta(t)}$ and using Eq.(\ref{defomegas}), one promptly shows that $\omega_s(t) = -\partial_t \theta(t)$.
By the linearity of the field amplitude with respect to the excited-state amplitude, the phase of the field is given by $\theta_b(t) = \theta(t)$, so that 
$\partial_t{\theta}_b(t) = \partial_t{\theta}(t)$.
Because $\phi^{(b)}(x_d,t)$ represents the quantum state of the electric field amplitude, the negative of the time derivative of its phase, 
$-\partial_t{\theta}_b(t)$, can be intuitively interpreted as an effective time-dependent color of the field $\omega^{(b)}_{\mathrm{eff}}(t) \equiv -\partial_t \theta_b(t)$.
Therefore, $\omega^{(b)}_{\mathrm{eff}}(t) = \omega_s(t)$, i.e., the instantaneous color of the field emitted through channel $b$ by the TLS is equal to the instantaneous transition frequency of the TLS itself.
Indeed, by testing this concept for the input packet $\phi^{(a)}(x-ct,0)$, we see that $\omega^{\mathrm{in}}_{\mathrm{eff}}(t) = \omega_L$ as we should expect.
In addition, the temporal modulation of $\omega_s(t)$, as described by Eq.\ref{wparticular}, is very slow, $\omega_s(t) - \omega_0 \sim \delta \ll \omega_0$, for the chosen parameters.
These two properties corroborate the interpretation of $\omega^{(b)}_{\mathrm{eff}}(t)$ as a time-dependent color for the reemitted photon through channel $b$.

Such an effective time-dependent color is more than a theoretical assessment.
This quantity can be evidenced experimentally.
In channel $a$, there can be interference between two amplitudes,
$\phi^{(a)}(x>0,t) = \phi^{(a)}(x-ct,0) + \sqrt{\beta}\ \psi(t-x/c)$.
The amplitude $\phi^{(a)}(x-ct,0)$ represents the free propagation of the input photon packet without interacting with the TLS, carrying its input effective frequency $\omega_L$.
The amplitude $\sqrt{\beta} \psi(t-x/c)$ represents the packet emitted through channel $a$ by the TLS dipole, that is symmetric with respect to the amplitude emitted through channel $b$, therefore carrying an effective frequency $\omega_s(t)$, as well.
Interference will depend on a frequency matching between these two frequencies.
To make that clear, we write the intensity on channel $a$, $I_a(t) = |\phi^{(a)}(|x_d|,t)|^2$ as an explicit function of 
$\omega_L$ and $\omega_s(t)$,
\begin{eqnarray}
&I_a(t) &= I_0(t) + I_b(t) +  \nn\\
          && + 2 \sqrt{I_0(t) I_b(t)} \cos{\left(\pi + \int_0^t [\omega_L - \omega_s(t')] \ dt'  \right)},
\label{Interference}
\end{eqnarray}
where $I_0(t) = |\phi^{(a)}(|x_d|-ct,0)|^2$ is the intensity of the input field.
We call to attention the fact that, in the monochromatic (long photon) regime, $\Delta \ll \Gamma_{\mathrm{1D}}$, we have that $I_b/I_0 = \Gamma_{\mathrm{1D}}^2/(\Gamma_{\mathrm{1D}}^2 + 4 \delta^2)$ and $I_a/I_0 = 4\delta^2/(\Gamma_{\mathrm{1D}}^2 + 4 \delta^2)$.
Therefore, at resonance, $\omega_s(t) = \omega_0 = \omega_L$, full reflection of the photon is recovered, $I_b/I_0 \rightarrow 1$ and $I_a/I_0 \rightarrow 0$, as predicted by other approaches to a similar scenario in the literature \cite{fan05, alexia, astafiev}.
Far-off-resonance, the photon is completely transmitted, $I_b/I_0 \rightarrow 0$ and $I_a/I_0 \rightarrow 1$, as expected.
The difference of the signals, $(I_a - I_0) - I_b$ is plotted in Fig.\ref{intensity}, as a function of time.
The parameters of the plots (a), (b) and (c) respectively correspond to the parameters of the curves black, blue and red from Fig.\ref{w}.
In all plots, the full lines represent the the case in which the correct time-dependent frequency $\omega_s(t)$ is taken into account, whereas the crosses represent the case where $\omega_s(t)$ is incorrectly approximated by the
static frequency $\omega_0$, showing a visible difference that can be measured from the field interference pattern.
This difference is much less pronounced in (a) and (b).
Plot (a) corresponds to the low detuned ($\delta = 0.1$) short packet ($\Delta = 5$).
The close-to-resonance condition imposes almost no shift, as evidenced by the coincidence between the curves.
Plot (b) corresponds to the highly detuned ($\delta = 5$) long packet ($\Delta = 0.1$).
In that case, the measurement of the theoretically predicted difference would demand very high experimental precision, since
$(I_a - I_0) - I_b \propto - 2\Delta \Gamma_\mathrm{1D}^2/(\Gamma_{\mathrm{1D}}^2+4\delta^2)$.
Plot (c) corresponds to the intermediate detuning ($\delta = 3$) and linewidth ($\Delta = 0.9$), in which case the contrast is most visible.

In conclusion, we have found a dynamical change in the TLS transition frequency induced by a propagating single-photon wave packet, which represents a realistic out-of-equilibrium initial quantum state of the field.
The time-dependence of the dynamics of the Stark shift is found as a function of the parameters of the packet, such as the linewidth and the central frequency.
The effect we evidence here is measurable in a field interference experiment.
A very timely perspective opened by this paper is the possibility of an autonomous control of the TLS frequency, without the need for external classical apparatuses. 
In Ref.\cite{inah13}, the authors experimentally control an artificial atom frequency by means of a strain-mediated coupling with a mechanically oscillating 1D waveguide.
We would like to investigate whether the dynamical Stark shift induced by a single-photon packet is able to drive oscillations in the mechanical degree of freedom of that hybrid optomechanical system, which could function as a storage for quantum work \cite{XuerebNJP}, harvested from the photon.




%
%
%

\begin{acknowledgements}
Conselho Nacional de Pesquisa CNPq Brazil (477612/2013-0, 478682/2013-1).
Instituto Nacional de Ci\^encia e Tecnologia -- Informa\c c\~ao Qu\^antica (INCT-IQ), Brazil.
Coordenação de Aperfeiçoamento de Pessoal de Nível Superior (CAPES) (88881.120135/2016-01).
\end{acknowledgements}

\end{document}